\documentclass[twocolumn,floatfix,superscriptaddress,aps,prr]{revtex4-1}
\usepackage{graphicx}
\usepackage{amsmath,mathtools}
\usepackage{braket}
\usepackage{bm}
\usepackage{booktabs}
\usepackage{hyperref}
\usepackage{pgfplots}
\usepackage{multirow}
\usepackage{amsfonts}
\usepackage{calc}
\usepackage{siunitx}
\usepackage{array}
\usepackage{color}
\usepackage[normalem]{ulem}

\usepackage{tikz}
\usepackage[utf8]{inputenc}
\usepackage{pgfplots}
\usepgfplotslibrary{groupplots,dateplot}
\usetikzlibrary{patterns,shapes.arrows}
\pgfplotsset{compat=newest}

\definecolor{ashgrey}{rgb}{0.7, 0.75, 0.71}
\definecolor{aurometalsaurus}{rgb}{0.43, 0.5, 0.5}

\hypersetup{
    colorlinks,
    linkcolor={red!50!black},
    citecolor={blue!50!black},
    urlcolor={blue!80!black}
}

\AtBeginDocument{
\heavyrulewidth=.08em
\lightrulewidth=.05em
\cmidrulewidth=.03em
\belowrulesep=.65ex
\belowbottomsep=0pt
\aboverulesep=.4ex
\abovetopsep=0pt
\cmidrulesep=\doublerulesep
\cmidrulekern=.5em
\defaultaddspace=.5em
}

\begin{document}
\title{Navigating the pitfalls of relic neutrino detection
}

\begin{abstract}{Beta-spectrum of radioactive atoms was long ago 
predicted to bear an imprint of the Cosmic Neutrino Background (C$\nu$B)~\cite{weinberg1962universal}. Over the years, it has been recognised that the best chance of achieving the signal-to-noise ratio required for the observation of this effect lies with solid-state designs~\cite{baracchini2018ptolemy}. Here we bring to the fore a fundamental quantum limitation on the type of beta-decayer that can be used in a such a design. We derive a simple usability criterion and show that $^3\rm H$, which is the most popular choice, fails to meet it. We provide a list of potentially suitable isotopes and discuss why their use in C$\nu$B detection requires further 
research.
} 
\end{abstract}

\author{Yevheniia Cheipesh}
\email{gene.cheypesh@gmail.com}
\affiliation{Instituut-Lorentz, Universiteit Leiden, P.O. Box 9506, 2300 RA Leiden, The Netherlands}

\author{Vadim Cheianov}
\affiliation{Instituut-Lorentz, Universiteit Leiden, P.O. Box 9506, 2300 RA Leiden, The Netherlands}

\author{Alexey Boyarsky}
\affiliation{Instituut-Lorentz, Universiteit Leiden, P.O. Box 9506, 2300 RA Leiden, The Netherlands}
\date{\today}

\maketitle

\textit{Introduction.}\,---\, The Cosmic Neutrino Background (C$\nu$B) is an unexplored source of precious cosmological data~\cite{weinberg1962universal}. Like the CMB, it carries a photographic image of 
the early Universe, albeit from a much older
epoch of neutrino decoupling. Although indirect evidence for the C$\nu$B was recently found in the Planck data~\cite{follin2015first}, direct detection of the relic neutrinos remains a major experimental challenge and a problem of great significance for the understanding of the pre-recombination age. The importance and basic principles of a C$\nu$B detection experiment were discussed as early as 1962 in a paper by S.~Weinberg~\cite{weinberg1962universal} who put forward the idea of a kinematical signature of 
the cosmic neutrino capture processes in beta-spectra of radioactive atoms. This idea was further elaborated in Ref.~\cite{Cocco:2007za}.

\begin{figure}[ht]
\includegraphics[scale=0.9]{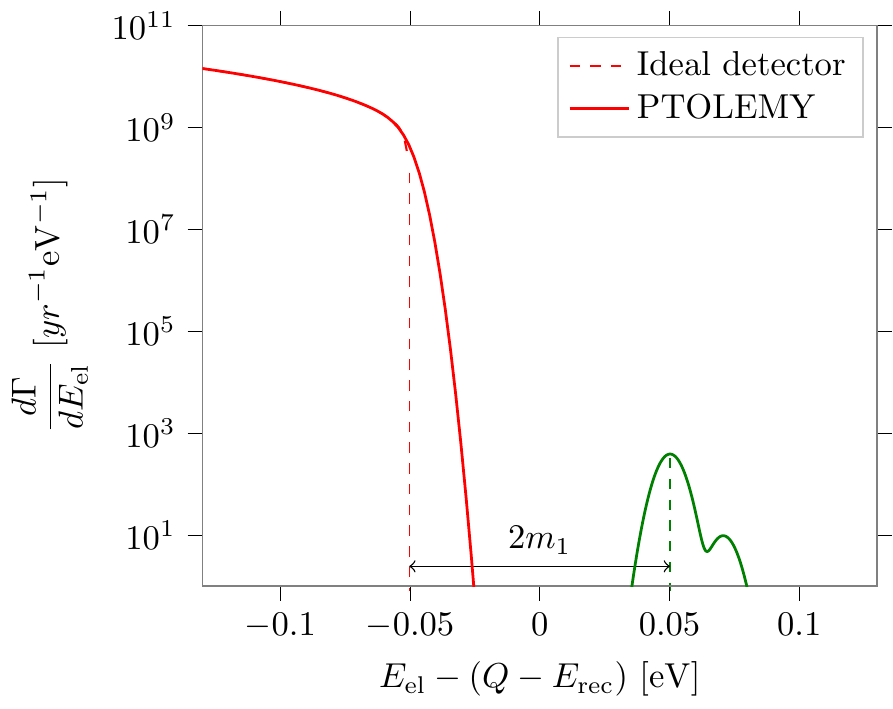}
\caption{\textbf{The $\beta$-spectrum of free monoatomic Tritium} centered around $Q - E_{\text{rec}}$, where $Q$ is the decay energy and $E_{\text{rec}}$ - recoil of the nucleus in the vacuum. The normal neutrino mass hierarchy~\cite{qian2015neutrino} is assumed with the mass of the lightest neutrino $m_1 = \SI{50}{\milli\electronvolt}$. The spontaneous $\beta$-decay spectrum is shown in red
while the C$\nu$B feature is shown in green.
The solid lines are drawn assuming a 
$10\, \rm meV$ resolution of the detector. 
}
\label{fig:spectrum}
\end{figure}

 The main roadblock in the way of the realisation of Weingerg's original proposal is the weakness of the neutrino-matter interaction, which makes it difficult to achieve a sufficient number of the relic neutrino capture events in a given radioactive sample. The problem is further compounded by the presence of a massive neutrino-emission background which imposes extremely stringent requirements on the 
energy resolution of the experiment~\cite{faessler2017can, cocco2009low}. The magnitude of the challenge is illustrated in FIG.~\ref{fig:spectrum} showing the $\beta$-emission spectrum of monoatomic $^3$H in vacuum. One can see that the spectrum  is dominated by the spontaneous $\beta$-decay background, shown in red, while the predicted signal~\cite{Betti:2019ouf} due to the relic neutrino capture process consists of a tiny feature shown in green~\footnote{The capture spectrum comprises of three peaks corresponding to the three neutrino mass eigenstates. The first two peaks overlap and are barely distinguishable.}. Not only is the predicted C$\nu$B feature 
quite weak, consisting of only a few events per year per $100$ g of $^3$H, but it is also positioned within a few tens of meV from the massive spontaneous decay background, which implies that the energy resolution of the experiment needs 
to be as good as $20 \rm \ meV.$ While the energy resolution specifications push the experimental apparatus towards a smaller scale, the extreme scarceness of useful events calls for a bigger working volume. The tension between these 
opposite requirements makes working with 
gaseous samples difficult, possibly impracticable.
The best to date experiment, KATRIN~\cite{wolf2010katrin}, which uses 
gaseous molecular Tritium as the working 
isotope falls short of the required sample 
activity by six orders of magnitude. It is worth noting that the sensitivity of experiments working 
with gaseous Tritium is further reduced due to
excitation of internal motions of the Tritium molecule and is further limited by the non-tritium background~\cite{bodine2015assessment, faessler2017can}.

Currently, the only viable alternative to the gas phase experiment is a solid state architecture where the $\beta$-emitters are adsorbed on a substrate~\cite{baracchini2018ptolemy}. Such a design can increase the event count by orders of magnitude while preserving the necessary degree of control over the emitted electrons. However, these advantages come at a price. 
In this paper we demonstrate that any solid state based $\beta$-decay experiment has fundamental limitations on its energy resolution, which are not related to the construction of the measuring apparatus. Such limitations arise from the quantum effect of the zero-point motion of the adsorbed $\beta$-emitter. We show that due to the 
extremely weak sensitivity of the zero-point motion to the details of the chemistry
of adsorption, the effect mainly imposes {\em intrinsic} requirements on the physical properties of the emitter~\footnote{In general, the interaction of an adsorbed radioactive atom with the substrate is complicated and it gives rise to several effects each contributing to the broadening of the measured $\beta$-emission spectrum.  In this paper, we only focus on one which is arguably the simplest and the strongest of all: the zero-point motion of an atom arising from the atom's adsorption.}.
In particular, we find that Tritium used in many existing and proposed experiments is not suitable for detecting C$\nu$B in a solid state setup. 
At the end, we list candidates for a suitable $\beta$-emitter and comment on what future theoretical and experimental research is needed to both confirm the choice of the atom and improve the 
resolution of the experiment.

\textit{Defining the problem}\,---\, 
 Although our analysis is not limited 
to a particular solid state design, we use 
for reference the setup of PTOLEMY~\cite{baracchini2018ptolemy}, a state of the art experimental proposal for the C$\nu$B detection that aims to achieve a sufficient number of events together with the required energy resolution of the apparatus~\cite{ betts2013development, Messina2018, Cocco2017nax, li2015detection, long2014detecting}. 
In PTOLEMY, mono atomic Tritium is deposited on graphene sheets arranged into a parallel stack and a clever magneto-electric design is used to extract 
and measure the energy of the electrons created 
in the two $\beta$-decay channels
\begin{align}\label{eq:neutrino_emission}
  \nonumber  \prescript{3}{}{\text{H}}  &\rightarrow \prescript{3}{}{\text{He}}  +  e + \bar{\nu}_e \\
 \nu_e +   \prescript{3}{}{\text{H}}  &\rightarrow \prescript{3}{}{\text{He}}  +  e 
\end{align}

\begin{figure*}[t]
\center
\includegraphics[scale=0.9]{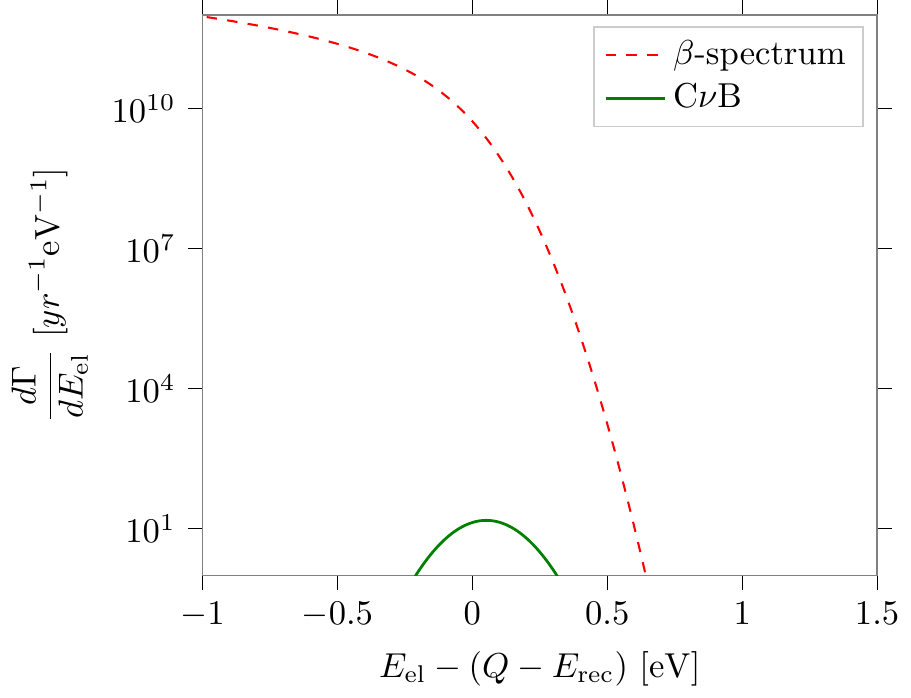}
\quad \includegraphics[scale=0.9]{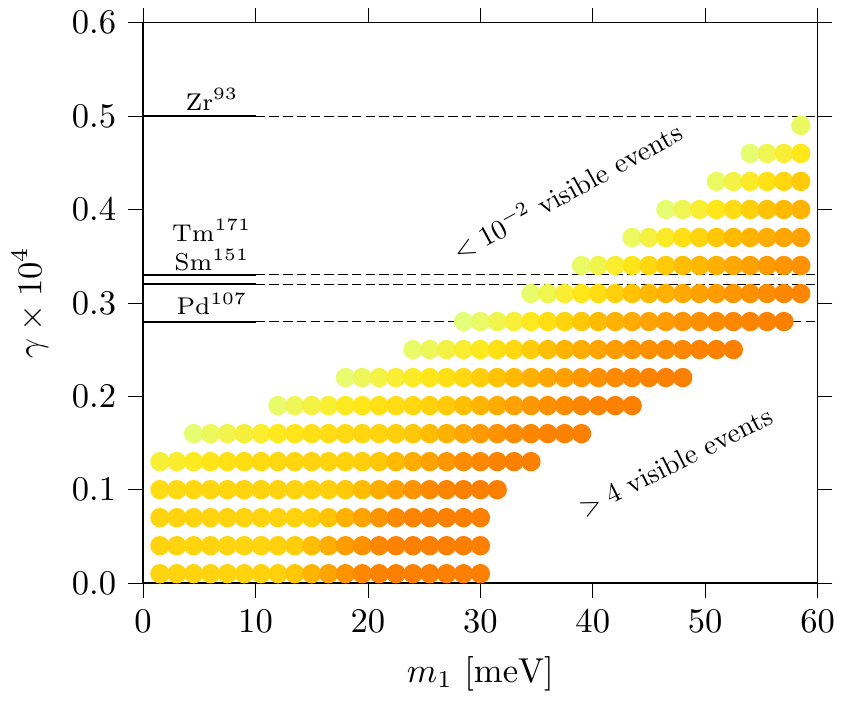} 
\caption{\textbf{The estimate of the smearing of the electron emission spectrum} due to the bonding of emitter to graphene. \textit{Left panel}: 
The electron emission spectrum for the  physisorbed atomic Tritium ($\lambda_{\text{osc}} = \SI{0.6}{\angstrom}$) taking the hierarchy, $m_1$ and energy resolution of the apparatus same as for the FIG.~\ref{fig:spectrum}. \textit{Right panel}: 
Visibility(defined by the number of C$\nu$B event that do not overlap with the continuous spectrum at all) of the C$\nu$B peak depending on the mass of the lightest neutrino $m_1$ and a dimensionless parameter $\gamma$ defined in Eq.~\eqref{eq:gammadef} that characterizes the emitter (for the physisorbed Tritium $\gamma \approx 3 \times 10^{-4}$).  The white areas on the bottom right and top left are correspondingly the areas of full and zero visibility and the coloured region in between corresponds to the partial visibility.
}\label{fig:spectrum_smeared}
\end{figure*}

The main goal of the C$\nu$B detection experiments is to detect the 
electrons produced in the neutrino capture channel (see FIG.~\ref{fig:spectrum}) that depends on the mass of the lightest neutrino and the hierarchy~\cite{mertens2015sensitivity, masood2007exact, Betti:2019ouf,  baracchini2018ptolemy}.  Since the captured relic neutrinos are  soft, it has a shape of $3$ narrow peaks~\footnote{Each of the peak corresponds to a separate mass eigenstate.} 
separated from the end of the main part of the spectrum by double the mass of the lightest neutrino. The spectrum depicted on FIG.~\ref{fig:spectrum} is calculated for an isolated Tritium atom
in the rest frame, where the recoil energy is  defined by the conservation laws.
However, if Tritium is absorbed on a substrate,
it can not be considered at rest and the recoil
energy  of the nucleus acquires some amount of uncertainty and so does the measured 
spectrum of the emitted electron (see FIG.~\ref{fig:spectrum_smeared}).

Two complementary views on such an uncertainty are possible, both leading to the 
same conclusion in the present context. In the ``semiclassical'' view the source 
of the uncertainty is the zero-point motion of the 
Tritium atom, which results in a fluctuating centre of mass 
frame at the moment of $\beta$-decay. In the fully quantum view the uncertainty 
results from quantum transitions of an atom into the highly excited vibrational 
states in the potential which confines it to the graphene sheet. We shall begin 
our discussion with the semi-classical picture.

It follows from Heisenberg uncertainty principle that 
an atom restricted to some finite region in space by the bonding potential cannot be exactly at 
rest. Even in the zero temperature limit it
performs a zero-point motion so that its velocity 
fluctuates randomly obeying some probability distribution 
$\mathcal{F}(\mathbf u).$ For localized states, $\mathcal{F}(\mathbf u)$ has a vanishing mean and dispersion defined by the Heisenberg uncertainty principle $\Delta u \sim \hbar/m_{\text{nucl}}\lambda_{\text{nucl}}$. Due to these random fluctuations in the velocity 
of the nucleus, the observed velocity distribution 
of the  emitted electron in the laboratory frame is given by the 
convolution 
\begin{equation}\label{eq:convolution}
    \tilde{\mathcal{G}}(\mathbf v) = 
    \int du \mathcal{F}(\mathbf u)\mathcal{G}(\mathbf v+ \mathbf u).
\end{equation}
where $\mathcal G(\mathbf v) $ is the velocity distribution of 
an electron emitted by a free Tritium atom at rest corresponding to the energy distribution given by a Fermi Golden Rule (see FIG.~\ref{fig:spectrum}). The formal applicability condition of Eq.~\eqref{eq:convolution} is that the spacing 
between the energy levels of the $^3\rm{He}^+$ ion emerging from $\beta$-decay be
much less than the typical recoil energy 
$\Delta \varepsilon  \ll E_{\text{rec}}.$ This condition is readily satisfied 
for the recoil energy in vacuum $E_{\text{rec}} = \SI{3.38}{\electronvolt}.$
We shall revisit this argument when we turn to the fully quantum picture. 

In the following analysis we will restrict ourselves to the particular case of the Tritium atoms adsorbed on the graphene following the PTOLEMY proposal. However the obtained results are also valid for more general bonding potentials (see the discussion at the end).

In the zero temperature limit, the function $\mathcal F(\mathbf u)$
appearing in Eq.~\eqref{eq:convolution}
is encoded in the wave function of the stationary state of a Tritium atom in the potential of the interaction of the atom with graphene.  Although such a potential has a rather complicated shape, 
as can be seen from multiple ab-initio studies~\cite{moaied2014theoretical, henwood2007ab, gonzalez2019hydrogen,ivanovskaya2010hydrogen},
the large mass of the nucleus justifies the use of the harmonic 
approximation near a local potential minimum
$$U = \frac{1}{2} \kappa_{i,j} r_i r_j  + U_0$$
where $r_i$ are the components of the atom's displacement 
vector and $\kappa$ is the Hessian tensor.
Then, it follows that $\mathcal F(\mathbf u)$ is a multivariate normal distribution
\begin{equation}\label{eq:F}
    \mathcal{F}(\mathbf u) = \dfrac{1}{(2\pi)^{3/2}}\dfrac{1}{\sqrt{\det\Sigma}}\exp\left(-\dfrac{1}{2} \sum_{i,j=1}^3 u_i\Sigma^{-1}_{i,j}u_j\right).
\end{equation}
with zero mean and a covariance matrix $\Sigma= \hbar m^{3/2} \sqrt \kappa.$ To find the latter, we proceed to the analysis of the bonding potential near its minima.

An adsorbed 
Tritium atom is predicted to occupy a symmetric position 
with respect to the graphene lattice, characterised by a C$_3$ point symmetry group. 
For this reason, the Hessian will generally have two distinct principal values, one corresponding 
to the axis orthogonal to graphene and one to 
the motion in the graphene plane  yielding two different potential profiles.

\begin{figure}[h!]
    \centering
\includegraphics[scale=1]{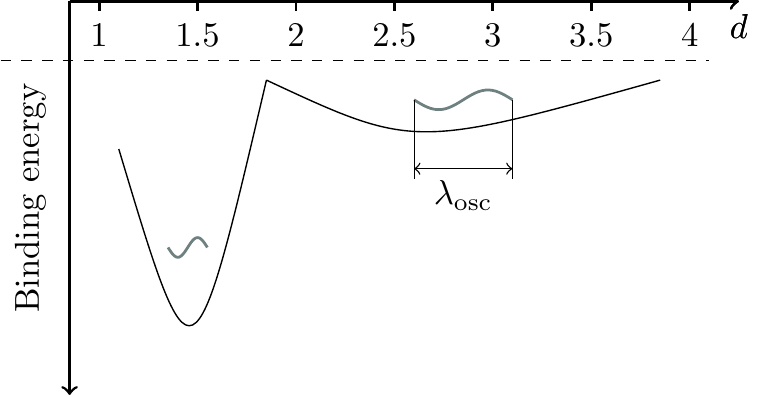}
    \caption{\textbf{Schematic profile of the potential} that bonds the Tritium atom in the direction perpendicular to the graphene. }
    \label{fig:potential_profile}
\end{figure}

According to the {\it ab initio} studies~\cite{ivanovskaya2010hydrogen,henwood2007ab,gonzalez2019hydrogen,moaied2014theoretical}, the potential that bonds the Tritium atom in the perpendicular direction has two minima, a  deep chemisorbtion minimum (in the range of $0.7 - \SI{3}{\electronvolt}$ for different studies) 
about $1.5 \rm \ \AA$  away from the graphene plane, and 
a shallow (about $\SI{0.2}{\electronvolt}$) physisorption minimum
$3 \rm \ \AA$ away from graphene~\footnote{We note, that we use the 
results of {\it ab initio} calculations for hydrogenated graphene. 
This is appropriate because Hydrogen is chemically equivalent to Tritium} (see FIG.~\ref{fig:potential_profile}).

The lateral motion of an atom is governed by the so-called migration potential~\cite{boukhvalov2010modeling}. The lateral stiffness in the case of chemisorption smaller than the vertical stiffness, however is substantial, as can be seen from Table~\ref{table:fit}. The case of a substrate producing a negligible 
migration potential will be discussed below.

 Introducing the normal displacement $z$ of an atom relative to the potential minimum, we can approximate the potential in the direction perpendicular to the graphene as $U(z) = \kappa z^2/2 + U_0.$  
The uncertainty in the position of 
the nucleus is then characterised 
by the oscillator length $\lambda^2= \hbar /\sqrt{m_{\rm nucl} \kappa}.$
The values 
of the constants $\kappa$ and $\lambda$ for different potential minima obtained from the fitting of the 
theoretical bonding profiles~\cite{moaied2014theoretical, henwood2007ab, gonzalez2019hydrogen, ivanovskaya2010hydrogen} are given in 
Table~\ref{table:fit}. The pronounced variability in the predicted values 
of the spring constant $\kappa$ is explained by the diversity of approximations 
used in different {\it ab initio} schemes. Note, however that the 
variability in the predicted values of the oscillator length is 
much less significant as  $\lambda \sim \kappa^{-\frac{1}{4}}$.
For this reason one can crudely neglect the difference between the strength of the lateral and normal confinement 
and consider the 
function $\mathcal F(\mathbf u)$ as approximately isotropic
\begin{equation}\label{eq:F1}
    \mathcal{F}(\mathbf u) \approx \dfrac{1}{\sqrt{2\pi }\Delta u} \exp\left(-\dfrac{1}{2} \dfrac{u^2}{\Delta u ^2}\right).
\end{equation}

We also note that, according to the Table~\ref{table:fit}, the typical predicted oscillator length 
is about an order of magnitude less than the typical length of the bond, which provides a posterior justification for the 
harmonic approximation. 
\begin{table}[t]
\centering
\begin{tabular}{c c ccc}
 \toprule
Potential & Source & $\kappa, \left[\SI{}{\electronvolt\per\angstrom^2}\right]$ & $\lambda, \left[\SI{}{\angstrom}\right]$  & $\Delta E, \left[\SI{}{\electronvolt}\right]$ \\
 \midrule
 & \cite{gonzalez2019hydrogen} & 2.15  & 0.16 & 0.60 \\
Chemisorption & \cite{moaied2014theoretical}, GGA & 4.62  & 0.13 &  0.73 \\
 & \cite{moaied2014theoretical}, vdW-DF & 4.9  & 0.13 & 0.75\\
 \midrule
\multirow{6}{*}{Physisorption}& \cite{ivanovskaya2010hydrogen} & 0.08  & 0.37 & 0.26\\
& \cite{gonzalez2019hydrogen} & 0.09  & 0.34  & 0.28 \\
& \cite{moaied2014theoretical}, GGA  & 0.18 & 0.29 & 0.33 \\
& \cite{moaied2014theoretical}, vdW-DF  & 0.13 & 0.32 & 0.3 \\
& \cite{henwood2007ab}, GGA  & 0.04 & 0.43 & 0.22\\
& \cite{henwood2007ab}, LDA  & 0.01 & 0.55 & 0.17\\
 \midrule
\\[-1em]
Migration & \cite{boukhvalov2010modeling} & 0.283  & 0.264 &  0.37\\
\bottomrule
\end{tabular}
\caption{\textbf{Harmonic fit} with the stiffness $\kappa$ of the chemisoption, physisorption potentials and the migration potential of the chemisorbed atom profiles near the minimum. $\lambda^2 = \hbar/\sqrt{m_{\text{nucl}}\kappa}$ and $\Delta E$ is the energy broadening of the emitted electron estimated from Eq.~\eqref{eq:var2}.}\label{table:fit}
\end{table}

 \textit{Estimate}\,---\, We are now in a position to obtain an estimate 
 for the uncertainty in the energy of an emitted electron. 
 By virtue of Heisenberg uncertainty principle, the variance of the velocity of the nucleus near a local potential minimum is 
 $\Delta u\approx \hbar /{m_{\text{nucl}} \lambda}.$
For an electron emitted at speed $v_{\text{el}}$ in the centre of mass frame 
the uncertainty of the energy measured in the laboratory frame is  
$\Delta E  \approx m_{\text{el}} v_{\text{el}}\Delta u,$ 
which near the edge of the 
electron emission spectrum can be written as 
\begin{equation}\label{eq:var2}
\Delta E \approx \frac{\hbar c}{\lambda_{\rm el} } \gamma
\end{equation}
where $\lambda^2_{\rm el} \equiv \hbar /\sqrt{m_{\rm el} \kappa}$ and we have introduced 
the dimensionless parameter 
\begin{equation}
\label{eq:gammadef}
\gamma = \left[ \frac{Q^2 m_{\rm el} }{m_{\text{nucl}}^3 c^4} \right]^{1/4},
\end{equation}
where $Q$  is the amount of energy released during the $\beta$ decay. Eqns.~\eqref{eq:var2},~\eqref{eq:gammadef} are the main result of this paper. 
This result, obtained so far 
 using semi-classical considerations, 
 can be cross-checked with a more precise 
 quantum mechanical calculation. For the latter, one applies the Fermi Golden Rule to the $\beta$-decay process where the initial state is the ground state of the atom in the harmonic potential and the final state is a product of neutrino, electron and atomic wave-functions that are highly excited WKB states (see  Appendix A for the detailed calculation). The result of such a calculation fully agrees with Eqns.~\eqref{eq:var2},~\eqref{eq:gammadef}.
 It is worth noting that in the fully quantum picture the  final $\beta$-spectrum in the C$\nu$B channel may  be continuous, discrete or mixed, depending on the depth of the bonding potential, but the overall envelope will be Gaussian with the width $\Delta E$.
 This is in agreement with  the previous results for the molecular Tritium~\cite{bodine2015assessment}~\footnote{As an example, the value of the stiffness $\kappa$ for the molecular tritium according to~\cite{bodine2015assessment} is $\kappa\approx \SI{75}{\electronvolt\per\angstrom^2}$. This is roughly $20$ times as large as the 
 corresponding value for the chemisorption (see Table~\ref{table:fit}). This means that the energy uncertainties $\Delta E$ in these two cases are of the same order which is in agreement with~\cite{bodine2015assessment}.}.

\textit{Discussion}\,---\,
In this paper, we have investigated the feasibility of the solid state based approach to the 
long-standing problem of detection of relic neutrino background. We conclude that, due to
 the remarkable progress in the 
 technology used for the measurement of electron emission spectrum (see e.g. ~\cite{baracchini2018ptolemy}) , the actual 
 energy resolution of the experiment is now controlled by a different bottleneck -  
  the uncertainties resulting from the interaction of the beta-emitter with the substrate. This paper addresses one type of such uncertainty considered -- the zero-point motion of the $\beta$-emitter. 
 For any given emitter it is practically 
 irreducible, which excludes certain emitters from 
 the list of suitable candidates for solid state setups.
In particular, for Tritium the uncertainty in the energy of the electrons
 is around $0.3-\SI{0.7}{\electronvolt}$ (see Table~\ref{table:fit} for the different bonding potentials according to different 
{\it ab initio} calculations), {\em i.e.} several times greater than the required energy resolution.

We see from Eqns.~\eqref{eq:var2}, \eqref{eq:gammadef} that the defining factor for  the energy uncertainty is the parameter $\gamma$ (see Eq.~\ref{eq:gammadef}), which only depends on the internal properties of a $\beta$-emitter
such as the mass of the nucleus and the energy released in the decay process. Therefore, a promising route to 
achieve a better performance of the detector would be to substitute a widely used Tritium~\cite{baracchini2018ptolemy, faessler2013search, wolf2010katrin, Cocco2017nax, betts2013development,Betti:2019ouf} with a heavier emitter (while simultaneously satisfying other experimental constraints, e.g. 
sufficiently long half-life time). The effect of the parameter $\gamma$ on the visibility of the C$\nu$B peak is shown on the right panel of FIG.~\ref{fig:spectrum_smeared}. One can see that, e.g., Tritium which has  $\gamma \approx 3\times 10^4,$ lies deep inside the region where the observation 
of the C$\nu$B peak is impossible. On the same figure we also indicate more suitable  $\beta$-emitters whose energy uncertainties are not prohibitive for the detection of the relic neutrinos with the masses $>\SI{20}{\milli\electronvolt}$.
 
Another important conclusion of our work is that 
although the energy uncertainty also depends on 
the bonding potential, this dependence only enters 
through the stiffness parameters and it is extremely weak $\Delta E \propto \kappa^{1/4}$. This implies that experimentation with different types of substrate is unlikely to make a substantial difference.  Indeed,  an order of magnitude improvement in $\Delta E,$ (which is needed for the state of the art experimental proposal~\cite{baracchini2018ptolemy}) would require a four orders of magnitude reduction in the value of $\kappa.$ Such a substantial deformation of the bonding potential presents a significant experimental challenge.

 A certain improvement in terms of the bonding potential  could still be achieved with adsorption that has a very weak lateral potential.
One such example is physisorption of Tritium on graphene. In the limiting case of a constant lateral potential, electrons emitted at grazing angles will not have any additional uncertainty to  their energy. Correspondingly, for the out-of-plane angles $\theta < \theta_{\text{max}} = \arcsin \left(\Delta E_{\text{max}}/\Delta E\right)$ the energy uncertainty will be bounded by $\Delta E_{\text{max}}$.  Here $\Delta E$ denotes the energy uncertainty for the isotropic case with finite mobility.  Restricting the detection collection  to $\theta < \theta_{\text{max}}$ reduces the number of events by a factor
 $\eta^{-1} \approx \pi \theta_{\text{max}}/90^{\circ}$. As an example, for $\Delta E_{\text{max}} = \SI{10}{\milli\electronvolt}$ one obtains $\theta_{\text{max}} \approx 3^\circ, \eta \approx 10$  which would entail the challenge of producing and handling 10 times as much radioactive material. This direction requires a full in-depth analysis which we leave for future studies.
  
We conclude, that a careful selection of the $\beta$-emitter (Fig.~\ref{fig:spectrum_smeared}) together with the use of an optimized substrate place C$\nu$B 
 detection potentially within the reach of the detection technologies developed by the PTOLEMY collaboration.

One should, however, note that the zero-point motion of the emitter does not exhaust the list of mechanisms that introduce uncertainty and errors into the beta-decay spectrum. Other potentially harmful mechanisms include the electrostatic interaction of the ionized atom with the substrate, charge relaxation in graphene, $X$-ray edge singularity, and phonon emission. We therefore strongly believe that further progress towards C$\nu$B detection requires a serious concerted effort both theoretical and experimental in the characterization of the physics and chemistry of the interaction of the $\beta$-emitter with its solid state environment.

\textit{Acknowledgements}\,---\,We are grateful to Chris Tully, A.P. Colijn and the whole PTOLEMY collaboration for fruitful discussions and feedback on the manuscript that allowed for its significant improvement. We also thank Kyrylo Bondarenko and Anastasiia Sokolenko for the useful discussion. YC is supported by the funding from the Netherlands Organization for Scientific Research (NWO/OCW) and from the European Research Council (ERC) under the European Union’s Horizon 2020 research and innovation programme. AB is supported by the European Research Council (ERC) Advanced Grant “NuBSM” (694896). VC is grateful to the Dutch Research Council (NWO) for partial support, grant No 680-91-130.

\begin{widetext}

\appendix

\section{ Quantum derivation of the energy uncertainty}

The aim of the fully quantum derivation is to underpin the semiclassical heuristic that was obtained in the main text as well as demonstrating its limitations. We note that we will not keep track of the pre-factors $\hbar, c$ and will restore them in the end. The rate of $\beta$-emission of an electron 
is given by the Fermi Golden Rule rule

\begin{equation}\label{eq:Gammadef}
\frac{d \Gamma}{dE} = \sum_{f}
2\pi\vert \braket{f|\hat V|i} \vert^2\delta(E_{i}- E_{f}) \delta(E-E_{f, \rm el}).
\end{equation} 
Here the vector $\vert i \rangle$ represents the initial state of the system having 
the energy $E_{\rm i}, $ the vector $\vert f \rangle,$ represents 
a final eigenstate of the Hamiltonian having the 
energy $E_{f}=E_{f, \rm el}+ E_{f, \rm He}$ where $E_{f, \rm el},$ is the 
kinetic energy of the outgoing electron and $E_{f, \rm He},$ is the energy of 
the $^3\rm He^+$ ion. The sum is performed over all such final states. 
The interaction potential $\hat V$ is responsible for $\beta$-decay 
vertex and is for our purposes an ultralocal product of the creation and 
annihilation operators of the fields involved in the process.

We make an assumption that the neutrino has zero kinetic energy. It is equivalent to restricting ourselves to region near the edge of the spectrum, which is exactly the region of interest to us. The energy conservation implies
\begin{equation}
\dfrac{\vec{k}^2}{2m_{\text{el}}} + \dfrac{\vec{p}^2}{2m_{\text{nucl}}} = \tilde{Q},
\end{equation} 
where  $\vec{k}$, $\vec{p}$ - are two-dimensional final momenta of the electron and nucleus respectively. $\tilde{Q}$ is the total energy of the nucleus before $\beta$-decay. 

The initial state of the system is a product of a plane wave state of an 
incoming relic neutrino, which it is safe to describe as a plane wave 
with nearly zero momentum, and the lowest energy eigenstate of a 
Tritium atom in the local minimum of the bonding potential.
As was discussed in the main text, such a state can be safely approximated 
as a ground state of a harmonic oscillator with two distinct principal stiffness eigenvalues (see table~\ref{table:fit}). The wave funcion of such 
a state has the form
\begin{equation}
\psi_{\text{i}}(\mathbf r ) \propto \exp\left(-\dfrac{z^2}{2\lambda_\perp^2}-\dfrac{\varrho^2}{2\lambda_\parallel^2}\right),
\end{equation}
where $z$ stands for the orthogonal displacement and $\varrho $ for the magnitude of the lateral displacement relative to the local potential minimum. Due to the in-plane symmetry of the graphene with respect to rotation, we can effectively restrict ourselves to a two-dimensional space $z, \varrho$.

The space of all possible final states $\ket{f}$ is quite large, and their wave functions may be quite complicated due to the intricate interaction of the $^3\rm He^+$ ion with
the graphene sheet. However, as we shall see momentarily the dominant contribution 
to the sum in \eqref{eq:Gammadef} comes from the states which are amenable 
to the WKB approximation and are therefore analytically tractable. Introducing the notation $\psi_f(\mathbf r)$ for the final state of the $^3\rm He^+$ ion, 
we write the matrix element in \eqref{eq:Gammadef} as 
\begin{equation}\label{eq:me}
 \bra{f}\hat V \ket{i} \sim \int d\mathbf r \psi^*_f(\mathbf r) \psi_i(\mathbf r) 
 e^{- i\mathbf k \mathbf r}
 \end{equation}
where $\mathbf k$ is the wave vector of the emitted electron at kinetic energy close 
to $Q.$ Since the electron's wave vector is quite large  $k \sim 10^2\, {\AA}^{-1}$
the rapid oscillations suppress the integral in Eq.~\eqref{eq:me} unless  
the state $\psi_f(\mathbf r)$ also contains an oscillatory factor, which 
has a roughly opposite De Broglie wave vector near $\mathbf r=0,$ where the support 
of $\psi_i(\mathbf r)$ is concentrated. This implies that the kinetic energy 
of the ion needs to be on the order of $3$eV, which exceeds the predicted chemisorption 
binding energy~\cite{ivanovskaya2010hydrogen,henwood2007ab,gonzalez2019hydrogen,moaied2014theoretical} and is orders of magnitude greater than the vibrational quantum near the potential minimum ($\hbar \omega \sim 0.01$ eV). Such highly excited states are 
generally characterised by a level spacing which is much narrower than the 
vibrational quantum near the minimum. 
They are also well described by semiclassical WKB wave functions, 
which on the scale of the oscillator length are indistinguishable from 
a plane wave.

With these considerations in mind, the application of the Fermi Golden Rule to such states gives 
 
\begin{equation}
\dfrac{d\Gamma}{dE}  \propto \left| \int_{-\infty}^{\infty} dx\int_{-\infty}^{\infty} dy\int_{-\infty}^{\infty} dz \exp\left(-i(k_x+p_x)x-i(k_y+p_y)y-i(k_z+p_z)z-\dfrac{x^2}{2\lambda^2_\parallel}-\dfrac{y^2}{2\lambda^2_\parallel}-\dfrac{z^2}{2\lambda^2_\perp}\right)\right|^2,
\end{equation} 
 where we have extended the integration over $z$ to $-\infty$. One can do it since the integrand is localized. $k/p_{x,y,z},$ are respectively the components of the electron and nucleus momenta that satisfy the energy conservation law
 \begin{align}
\nonumber |p| & = \sqrt{2m_{\text{nucl}}\left(\tilde{Q} - E_{\text{el}}\right)}\\
|k| & = \sqrt{2m_{\text{el}} E_{\text{el}}}  
\end{align}
We re-scale coordinates $\tilde{r}_i = \dfrac{r_i}{\sqrt{2}\lambda_i}$ and obtain
\begin{equation}
\dfrac{d\Gamma}{dE} \propto \left|\int_{-\infty}^{\infty} d\tilde{x}\int_{-\infty}^{\infty} d\tilde{y}\int_{-\infty}^{\infty} d\tilde{z} \exp\left(-i\sqrt{2}\lambda_\parallel (k_x+p_x)-i\sqrt{2}\lambda_\parallel (k_y+p_y)-i\sqrt{2}\lambda_\perp (k_\perp+p_\perp)\tilde{z}-\tilde{x}^2-\tilde{y}^2-\tilde{z}^2\right)\right|^2,
\end{equation} 
that can be brought to a Gauss integral
\begin{align}\label{eq:int1}
\nonumber \dfrac{d\Gamma}{dE}& \propto  e^{-\lambda^2_\perp(k_\perp+p_\perp)^2-\lambda^2_\parallel(k_\parallel+p_\parallel)^2} \times \\
&\left| \int_{-\infty}^{\infty} d\tilde{x}\int_{-\infty}^{\infty} d\tilde{y}\int_{-\infty}^{\infty} d\tilde{z} \exp\left(-\left(\tilde{x} + \dfrac{i\lambda_\parallel(k_x + p_x)}{\sqrt{2}}\right)^2 -\left(\tilde{y} + \dfrac{i\lambda_\parallel(k_y + p_y)}{\sqrt{2}}\right)^2 -\left(\tilde{z} + \dfrac{i\lambda_\perp(k_z + p_z)}{\sqrt{2}}\right)^2\right)\right|^2,
\end{align} 
where $k_\parallel/p_\parallel^2 = k_x/p_x^2 + k_y/p_y^2, p_\perp/p_\perp = k_z/p_z$. Integrating Eq.~\ref{eq:int1} gives the Gaussian distribution
\begin{equation}\label{eq:main}
\dfrac{d\Gamma}{dE} \propto e^{-\lambda^2_\perp(k_\perp+p_\perp)^2-\lambda^2_\parallel(k_\parallel+p_\parallel)^2}.
\end{equation} 

\begin{figure*}[t]
\center
\includegraphics[scale=0.9]{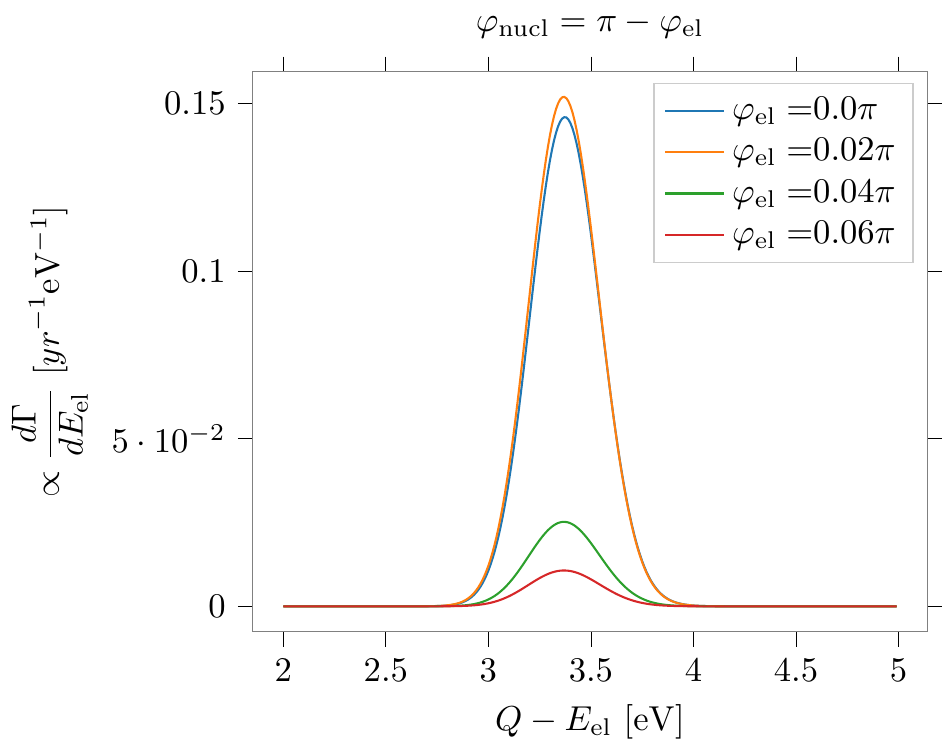} \quad
\includegraphics[scale=0.9]{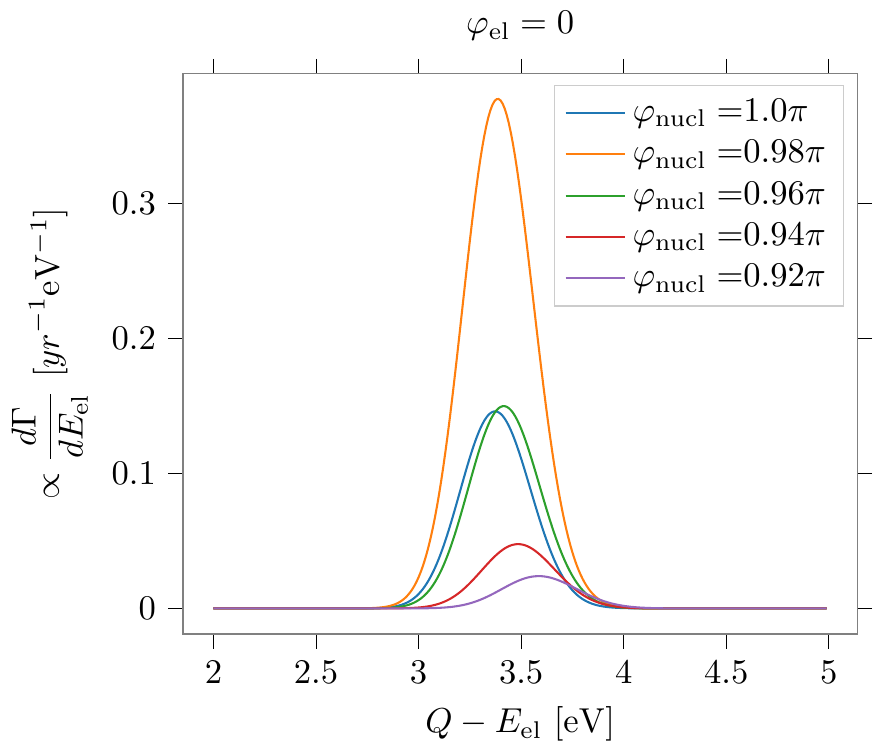}
\caption{\textbf{Distribution function} (not normalized) of the energy of the electron near the edge of the spectrum. Electron and nucleus are emitted with the corresponding angles $\varphi_{e/\text{nucl}}$ (relative to the axes perpendicular to the graphene substrate).}\label{fig:variance}
\end{figure*}
The distribution Eq.~\eqref{eq:main} depends on the angles of the emitted nucleus and electron $\varphi_{1,2}$. These angles are taken relative to the axes perpendicular to the graphene substrate.
\begin{equation}
\dfrac{d\Gamma}{dE} \propto e^{-\lambda^2_\perp\left(|k|\cos\varphi_2 + |p|\cos\varphi_1\right)^2-\lambda^2_\parallel\left(|k|\sin\varphi_2 + |p|\sin\varphi_1\right)^2},
\end{equation} 
Let us estimate the variance of this distribution for the normal emission of the electron
\begin{equation}
\dfrac{d\Gamma}{dE} \propto e^{-\lambda^2\left(k -p\right)^2},
\end{equation} 
where $k = \sqrt{2m_{\text{el}} E_{\text{el}}}, p =  \sqrt{2m_{\text{nucl}}\left(\tilde{Q}- E_{\text{el}}\right)}$.

In order to obtain the variance, wee need to expand near the maximum of the distribution that corresponds to its mean. If we write everything in terms of the deviation from the mean energy of the electron $\delta E_{\text{el}} = \tilde{Q} - E_{\text{rec}} - E_{\text{el}}$ 
\begin{align}
 \nonumber k &= \sqrt{2m_{\text{el}} \left(\tilde{Q} - E_{\text{rec}} - \delta E_{\text{el}}\right)} \approx \sqrt{2 m_{\text{el}} \left(\tilde{Q} - E_{\text{rec}}\right)}\left(1  - \dfrac{\delta E_{\text{el}}}{2(\tilde{Q} - E_{\text{rec}})}\right)  \\
 p &= \sqrt{2 m_{\text{nucl}} \left(E_{\text{rec}} + \delta E_{\text{el}}\right)}\approx \sqrt{2 m_{\text{nucl}} E_{\text{rec}}}\left(1  + \dfrac{\delta E_{\text{el}}}{2 E_{\text{rec}}}\right) .
\end{align}
Accounting to the fact that $E_{\text{rec}} \approx \dfrac{m_{\text{el}}}{m_{\text{nucl}}} \tilde{Q}$,
\begin{align}
 \nonumber k & \approx \sqrt{2 m_{\text{el}} \tilde{Q}}\left(1  - \dfrac{\delta E_{\text{el}}}{2\tilde{Q}}\right)  \\
 p &\approx \sqrt{2 m_{\text{el}} \tilde{Q}}\left(1  + \dfrac{m_{\text{nucl}}}{m_{\text{el}}}\dfrac{\delta E_{\text{el}}}{2 \tilde{Q}}\right) .
\end{align}
With this we obtain Gaussian distribution
\begin{align}
\nonumber \dfrac{d\Gamma}{dE} \propto \exp&\left(-\dfrac{\lambda^2 m_{\text{nucl}}^2}{2m_{\text{el}} \tilde{Q}}\delta E_{\text{el}}^2 \right),
\end{align}
with the variance with the restored units is
\begin{equation}
    \sigma = \dfrac{\hbar}{\lambda}\dfrac{\sqrt{\tilde{Q} m_{\text{el}}}}{ m_{\text{nucl}} }.
\end{equation}
\end{widetext}

\bibliographystyle{apsrev4-1}
\bibliography{chem_refs}

\begin{thebibliography}{28}%
\makeatletter
\providecommand \@ifxundefined [1]{%
 \@ifx{#1\undefined}
}%
\providecommand \@ifnum [1]{%
 \ifnum #1\expandafter \@firstoftwo
 \else \expandafter \@secondoftwo
 \fi
}%
\providecommand \@ifx [1]{%
 \ifx #1\expandafter \@firstoftwo
 \else \expandafter \@secondoftwo
 \fi
}%
\providecommand \natexlab [1]{#1}%
\providecommand \enquote  [1]{``#1''}%
\providecommand \bibnamefont  [1]{#1}%
\providecommand \bibfnamefont [1]{#1}%
\providecommand \citenamefont [1]{#1}%
\providecommand \href@noop [0]{\@secondoftwo}%
\providecommand \href [0]{\begingroup \@sanitize@url \@href}%
\providecommand \@href[1]{\@@startlink{#1}\@@href}%
\providecommand \@@href[1]{\endgroup#1\@@endlink}%
\providecommand \@sanitize@url [0]{\catcode `\\12\catcode `\$12\catcode
  `\&12\catcode `\#12\catcode `\^12\catcode `\_12\catcode `\%12\relax}%
\providecommand \@@startlink[1]{}%
\providecommand \@@endlink[0]{}%
\providecommand \url  [0]{\begingroup\@sanitize@url \@url }%
\providecommand \@url [1]{\endgroup\@href {#1}{\urlprefix }}%
\providecommand \urlprefix  [0]{URL }%
\providecommand \Eprint [0]{\href }%
\providecommand \doibase [0]{http://dx.doi.org/}%
\providecommand \selectlanguage [0]{\@gobble}%
\providecommand \bibinfo  [0]{\@secondoftwo}%
\providecommand \bibfield  [0]{\@secondoftwo}%
\providecommand \translation [1]{[#1]}%
\providecommand \BibitemOpen [0]{}%
\providecommand \bibitemStop [0]{}%
\providecommand \bibitemNoStop [0]{.\EOS\space}%
\providecommand \EOS [0]{\spacefactor3000\relax}%
\providecommand \BibitemShut  [1]{\csname bibitem#1\endcsname}%
\let\auto@bib@innerbib\@empty
\bibitem [{\citenamefont {Weinberg}(1962)}]{weinberg1962universal}%
  \BibitemOpen
  \bibfield  {author} {\bibinfo {author} {\bibfnamefont {S.}~\bibnamefont
  {Weinberg}},\ }\href@noop {} {\bibfield  {journal} {\bibinfo  {journal}
  {Physical Review}\ }\textbf {\bibinfo {volume} {128}},\ \bibinfo {pages}
  {1457} (\bibinfo {year} {1962})}\BibitemShut {NoStop}%
\bibitem [{\citenamefont {Baracchini}\ \emph {et~al.}(2018)\citenamefont
  {Baracchini}, \citenamefont {Betti}, \citenamefont {Biasotti}, \citenamefont
  {Bosca}, \citenamefont {Calle}, \citenamefont {Carabe-Lopez}, \citenamefont
  {Cavoto}, \citenamefont {Chang}, \citenamefont {Cocco}, \citenamefont
  {Colijn} \emph {et~al.}}]{baracchini2018ptolemy}%
  \BibitemOpen
  \bibfield  {author} {\bibinfo {author} {\bibfnamefont {E.}~\bibnamefont
  {Baracchini}}, \bibinfo {author} {\bibfnamefont {M.}~\bibnamefont {Betti}},
  \bibinfo {author} {\bibfnamefont {M.}~\bibnamefont {Biasotti}}, \bibinfo
  {author} {\bibfnamefont {A.}~\bibnamefont {Bosca}}, \bibinfo {author}
  {\bibfnamefont {F.}~\bibnamefont {Calle}}, \bibinfo {author} {\bibfnamefont
  {J.}~\bibnamefont {Carabe-Lopez}}, \bibinfo {author} {\bibfnamefont
  {G.}~\bibnamefont {Cavoto}}, \bibinfo {author} {\bibfnamefont
  {C.}~\bibnamefont {Chang}}, \bibinfo {author} {\bibfnamefont
  {A.}~\bibnamefont {Cocco}}, \bibinfo {author} {\bibfnamefont
  {A.}~\bibnamefont {Colijn}},  \emph {et~al.},\ }\href@noop {} {\bibfield
  {journal} {\bibinfo  {journal} {arXiv preprint arXiv:1808.01892}\ } (\bibinfo
  {year} {2018})}\BibitemShut {NoStop}%
\bibitem [{\citenamefont {Follin}\ \emph {et~al.}(2015)\citenamefont {Follin},
  \citenamefont {Knox}, \citenamefont {Millea},\ and\ \citenamefont
  {Pan}}]{follin2015first}%
  \BibitemOpen
  \bibfield  {author} {\bibinfo {author} {\bibfnamefont {B.}~\bibnamefont
  {Follin}}, \bibinfo {author} {\bibfnamefont {L.}~\bibnamefont {Knox}},
  \bibinfo {author} {\bibfnamefont {M.}~\bibnamefont {Millea}}, \ and\ \bibinfo
  {author} {\bibfnamefont {Z.}~\bibnamefont {Pan}},\ }\href@noop {} {\bibfield
  {journal} {\bibinfo  {journal} {Physical review letters}\ }\textbf {\bibinfo
  {volume} {115}},\ \bibinfo {pages} {091301} (\bibinfo {year}
  {2015})}\BibitemShut {NoStop}%
\bibitem [{\citenamefont {Cocco}\ \emph {et~al.}(2007)\citenamefont {Cocco},
  \citenamefont {Mangano},\ and\ \citenamefont {Messina}}]{Cocco:2007za}%
  \BibitemOpen
  \bibfield  {author} {\bibinfo {author} {\bibfnamefont {A.~G.}\ \bibnamefont
  {Cocco}}, \bibinfo {author} {\bibfnamefont {G.}~\bibnamefont {Mangano}}, \
  and\ \bibinfo {author} {\bibfnamefont {M.}~\bibnamefont {Messina}},\ }\href
  {\doibase 10.1088/1475-7516/2007/06/015} {\bibfield  {journal} {\bibinfo
  {journal} {JCAP}\ }\textbf {\bibinfo {volume} {06}},\ \bibinfo {pages} {015}
  (\bibinfo {year} {2007})},\ \Eprint {http://arxiv.org/abs/hep-ph/0703075}
  {arXiv:hep-ph/0703075} \BibitemShut {NoStop}%
\bibitem [{\citenamefont {Faessler}\ \emph {et~al.}(2017)\citenamefont
  {Faessler}, \citenamefont {Hod{\'a}k}, \citenamefont {Kovalenko},\ and\
  \citenamefont {{\v{S}}imkovic}}]{faessler2017can}%
  \BibitemOpen
  \bibfield  {author} {\bibinfo {author} {\bibfnamefont {A.}~\bibnamefont
  {Faessler}}, \bibinfo {author} {\bibfnamefont {R.}~\bibnamefont {Hod{\'a}k}},
  \bibinfo {author} {\bibfnamefont {S.}~\bibnamefont {Kovalenko}}, \ and\
  \bibinfo {author} {\bibfnamefont {F.}~\bibnamefont {{\v{S}}imkovic}},\ }in\
  \href@noop {} {\emph {\bibinfo {booktitle} {Quarks, Nuclei and Stars:
  Memorial Volume Dedicated to Gerald E. Brown}}}\ (\bibinfo  {publisher}
  {World Scientific},\ \bibinfo {year} {2017})\ pp.\ \bibinfo {pages}
  {81--91}\BibitemShut {NoStop}%
\bibitem [{\citenamefont {Cocco}\ \emph {et~al.}(2009)\citenamefont {Cocco},
  \citenamefont {Mangano},\ and\ \citenamefont {Messina}}]{cocco2009low}%
  \BibitemOpen
  \bibfield  {author} {\bibinfo {author} {\bibfnamefont {A.~G.}\ \bibnamefont
  {Cocco}}, \bibinfo {author} {\bibfnamefont {G.}~\bibnamefont {Mangano}}, \
  and\ \bibinfo {author} {\bibfnamefont {M.}~\bibnamefont {Messina}},\
  }\href@noop {} {\bibfield  {journal} {\bibinfo  {journal} {Physical Review
  D}\ }\textbf {\bibinfo {volume} {79}},\ \bibinfo {pages} {053009} (\bibinfo
  {year} {2009})}\BibitemShut {NoStop}%
\bibitem [{\citenamefont {Betti}\ \emph {et~al.}(2019)\citenamefont {Betti}
  \emph {et~al.}}]{Betti:2019ouf}%
  \BibitemOpen
  \bibfield  {author} {\bibinfo {author} {\bibfnamefont {M.}~\bibnamefont
  {Betti}} \emph {et~al.} (\bibinfo {collaboration} {PTOLEMY}),\ }\href
  {\doibase 10.1088/1475-7516/2019/07/047} {\bibfield  {journal} {\bibinfo
  {journal} {JCAP}\ }\textbf {\bibinfo {volume} {07}},\ \bibinfo {pages} {047}
  (\bibinfo {year} {2019})},\ \Eprint {http://arxiv.org/abs/1902.05508}
  {arXiv:1902.05508 [astro-ph.CO]} \BibitemShut {NoStop}%
\bibitem [{Note1()}]{Note1}%
  \BibitemOpen
  \bibinfo {note} {The capture spectrum comprises of three peaks corresponding
  to the three neutrino mass eigenstates. The first two peaks overlap and are
  barely distinguishable.}\BibitemShut {Stop}%
\bibitem [{\citenamefont {Wolf}\ \emph {et~al.}(2010)\citenamefont {Wolf},
  \citenamefont {Collaboration} \emph {et~al.}}]{wolf2010katrin}%
  \BibitemOpen
  \bibfield  {author} {\bibinfo {author} {\bibfnamefont {J.}~\bibnamefont
  {Wolf}}, \bibinfo {author} {\bibfnamefont {K.}~\bibnamefont {Collaboration}},
   \emph {et~al.},\ }\href@noop {} {\bibfield  {journal} {\bibinfo  {journal}
  {Nuclear Instruments and Methods in Physics Research Section A: Accelerators,
  Spectrometers, Detectors and Associated Equipment}\ }\textbf {\bibinfo
  {volume} {623}},\ \bibinfo {pages} {442} (\bibinfo {year}
  {2010})}\BibitemShut {NoStop}%
\bibitem [{\citenamefont {Bodine}\ \emph {et~al.}(2015)\citenamefont {Bodine},
  \citenamefont {Parno},\ and\ \citenamefont
  {Robertson}}]{bodine2015assessment}%
  \BibitemOpen
  \bibfield  {author} {\bibinfo {author} {\bibfnamefont {L.}~\bibnamefont
  {Bodine}}, \bibinfo {author} {\bibfnamefont {D.}~\bibnamefont {Parno}}, \
  and\ \bibinfo {author} {\bibfnamefont {R.}~\bibnamefont {Robertson}},\
  }\href@noop {} {\bibfield  {journal} {\bibinfo  {journal} {Physical Review
  C}\ }\textbf {\bibinfo {volume} {91}},\ \bibinfo {pages} {035505} (\bibinfo
  {year} {2015})}\BibitemShut {NoStop}%
\bibitem [{Note2()}]{Note2}%
  \BibitemOpen
  \bibinfo {note} {In general, the interaction of an adsorbed radioactive atom
  with the substrate is complicated and it gives rise to several effects each
  contributing to the broadening of the measured $\beta $-emission spectrum. In
  this paper, we only focus on one which is arguably the simplest and the
  strongest of all: the zero-point motion of an atom arising from the atom's
  adsorption.}\BibitemShut {Stop}%
\bibitem [{\citenamefont {Betts}\ \emph {et~al.}(2013)\citenamefont {Betts},
  \citenamefont {Blanchard}, \citenamefont {Carnevale}, \citenamefont {Chang},
  \citenamefont {Chen}, \citenamefont {Chidzik}, \citenamefont {Ciebiera},
  \citenamefont {Cloessner}, \citenamefont {Cocco}, \citenamefont {Cohen} \emph
  {et~al.}}]{betts2013development}%
  \BibitemOpen
  \bibfield  {author} {\bibinfo {author} {\bibfnamefont {S.}~\bibnamefont
  {Betts}}, \bibinfo {author} {\bibfnamefont {W.}~\bibnamefont {Blanchard}},
  \bibinfo {author} {\bibfnamefont {R.}~\bibnamefont {Carnevale}}, \bibinfo
  {author} {\bibfnamefont {C.}~\bibnamefont {Chang}}, \bibinfo {author}
  {\bibfnamefont {C.}~\bibnamefont {Chen}}, \bibinfo {author} {\bibfnamefont
  {S.}~\bibnamefont {Chidzik}}, \bibinfo {author} {\bibfnamefont
  {L.}~\bibnamefont {Ciebiera}}, \bibinfo {author} {\bibfnamefont
  {P.}~\bibnamefont {Cloessner}}, \bibinfo {author} {\bibfnamefont
  {A.}~\bibnamefont {Cocco}}, \bibinfo {author} {\bibfnamefont
  {A.}~\bibnamefont {Cohen}},  \emph {et~al.},\ }\href@noop {} {\bibfield
  {journal} {\bibinfo  {journal} {arXiv preprint arXiv:1307.4738}\ } (\bibinfo
  {year} {2013})}\BibitemShut {NoStop}%
\bibitem [{\citenamefont {Messina}(2018)}]{Messina2018}%
  \BibitemOpen
  \bibfield  {author} {\bibinfo {author} {\bibfnamefont {M.}~\bibnamefont
  {Messina}},\ }\href@noop {} {\bibfield  {journal} {\bibinfo  {journal}
  {Frascati Phys. Ser.}\ } (\bibinfo {year} {2018})}\BibitemShut {NoStop}%
\bibitem [{\citenamefont {Cocco}(2017)}]{Cocco2017nax}%
  \BibitemOpen
  \bibfield  {author} {\bibinfo {author} {\bibfnamefont {A.~G.}\ \bibnamefont
  {Cocco}},\ }\href {\doibase 10.22323/1.283.0092} {\bibfield  {journal}
  {\bibinfo  {journal} {PoS}\ } (\bibinfo {year} {2017}),\
  10.22323/1.283.0092}\BibitemShut {NoStop}%
\bibitem [{\citenamefont {Li}(2015)}]{li2015detection}%
  \BibitemOpen
  \bibfield  {author} {\bibinfo {author} {\bibfnamefont {Y.-F.}\ \bibnamefont
  {Li}},\ }\href@noop {} {\bibfield  {journal} {\bibinfo  {journal}
  {International Journal of Modern Physics A}\ }\textbf {\bibinfo {volume}
  {30}},\ \bibinfo {pages} {1530031} (\bibinfo {year} {2015})}\BibitemShut
  {NoStop}%
\bibitem [{\citenamefont {Long}\ \emph {et~al.}(2014)\citenamefont {Long},
  \citenamefont {Lunardini},\ and\ \citenamefont
  {Sabancilar}}]{long2014detecting}%
  \BibitemOpen
  \bibfield  {author} {\bibinfo {author} {\bibfnamefont {A.~J.}\ \bibnamefont
  {Long}}, \bibinfo {author} {\bibfnamefont {C.}~\bibnamefont {Lunardini}}, \
  and\ \bibinfo {author} {\bibfnamefont {E.}~\bibnamefont {Sabancilar}},\
  }\href@noop {} {\bibfield  {journal} {\bibinfo  {journal} {Journal of
  Cosmology and Astroparticle Physics}\ }\textbf {\bibinfo {volume} {2014}},\
  \bibinfo {pages} {038} (\bibinfo {year} {2014})}\BibitemShut {NoStop}%
\bibitem [{\citenamefont {Qian}\ and\ \citenamefont
  {Vogel}(2015)}]{qian2015neutrino}%
  \BibitemOpen
  \bibfield  {author} {\bibinfo {author} {\bibfnamefont {X.}~\bibnamefont
  {Qian}}\ and\ \bibinfo {author} {\bibfnamefont {P.}~\bibnamefont {Vogel}},\
  }\href@noop {} {\bibfield  {journal} {\bibinfo  {journal} {Progress in
  Particle and Nuclear Physics}\ }\textbf {\bibinfo {volume} {83}},\ \bibinfo
  {pages} {1} (\bibinfo {year} {2015})}\BibitemShut {NoStop}%
\bibitem [{\citenamefont {Mertens}\ \emph {et~al.}(2015)\citenamefont
  {Mertens}, \citenamefont {Lasserre}, \citenamefont {Groh}, \citenamefont
  {Drexlin}, \citenamefont {Glueck}, \citenamefont {Huber}, \citenamefont
  {Poon}, \citenamefont {Steidl}, \citenamefont {Steinbrink},\ and\
  \citenamefont {Weinheimer}}]{mertens2015sensitivity}%
  \BibitemOpen
  \bibfield  {author} {\bibinfo {author} {\bibfnamefont {S.}~\bibnamefont
  {Mertens}}, \bibinfo {author} {\bibfnamefont {T.}~\bibnamefont {Lasserre}},
  \bibinfo {author} {\bibfnamefont {S.}~\bibnamefont {Groh}}, \bibinfo {author}
  {\bibfnamefont {G.}~\bibnamefont {Drexlin}}, \bibinfo {author} {\bibfnamefont
  {F.}~\bibnamefont {Glueck}}, \bibinfo {author} {\bibfnamefont
  {A.}~\bibnamefont {Huber}}, \bibinfo {author} {\bibfnamefont
  {A.}~\bibnamefont {Poon}}, \bibinfo {author} {\bibfnamefont {M.}~\bibnamefont
  {Steidl}}, \bibinfo {author} {\bibfnamefont {N.}~\bibnamefont {Steinbrink}},
  \ and\ \bibinfo {author} {\bibfnamefont {C.}~\bibnamefont {Weinheimer}},\
  }\href@noop {} {\bibfield  {journal} {\bibinfo  {journal} {Journal of
  Cosmology and Astroparticle Physics}\ }\textbf {\bibinfo {volume} {2015}},\
  \bibinfo {pages} {020} (\bibinfo {year} {2015})}\BibitemShut {NoStop}%
\bibitem [{\citenamefont {Masood}\ \emph {et~al.}(2007)\citenamefont {Masood},
  \citenamefont {Nasri}, \citenamefont {Schechter}, \citenamefont
  {T{\'o}rtola}, \citenamefont {Valle},\ and\ \citenamefont
  {Weinheimer}}]{masood2007exact}%
  \BibitemOpen
  \bibfield  {author} {\bibinfo {author} {\bibfnamefont {S.~S.}\ \bibnamefont
  {Masood}}, \bibinfo {author} {\bibfnamefont {S.}~\bibnamefont {Nasri}},
  \bibinfo {author} {\bibfnamefont {J.}~\bibnamefont {Schechter}}, \bibinfo
  {author} {\bibfnamefont {M.~A.}\ \bibnamefont {T{\'o}rtola}}, \bibinfo
  {author} {\bibfnamefont {J.~W.}\ \bibnamefont {Valle}}, \ and\ \bibinfo
  {author} {\bibfnamefont {C.}~\bibnamefont {Weinheimer}},\ }\href@noop {}
  {\bibfield  {journal} {\bibinfo  {journal} {Physical Review C}\ }\textbf
  {\bibinfo {volume} {76}},\ \bibinfo {pages} {045501} (\bibinfo {year}
  {2007})}\BibitemShut {NoStop}%
\bibitem [{Note3()}]{Note3}%
  \BibitemOpen
  \bibinfo {note} {Each of the peak corresponds to a separate mass
  eigenstate.}\BibitemShut {Stop}%
\bibitem [{\citenamefont {Moaied}\ \emph {et~al.}(2014)\citenamefont {Moaied},
  \citenamefont {Moreno}, \citenamefont {Caturla}, \citenamefont
  {Yndur{\'a}in},\ and\ \citenamefont {Palacios}}]{moaied2014theoretical}%
  \BibitemOpen
  \bibfield  {author} {\bibinfo {author} {\bibfnamefont {M.}~\bibnamefont
  {Moaied}}, \bibinfo {author} {\bibfnamefont {J.}~\bibnamefont {Moreno}},
  \bibinfo {author} {\bibfnamefont {M.}~\bibnamefont {Caturla}}, \bibinfo
  {author} {\bibfnamefont {F.}~\bibnamefont {Yndur{\'a}in}}, \ and\ \bibinfo
  {author} {\bibfnamefont {J.}~\bibnamefont {Palacios}},\ }\href@noop {}
  {\bibfield  {journal} {\bibinfo  {journal} {arXiv preprint arXiv:1405.3165}\
  } (\bibinfo {year} {2014})}\BibitemShut {NoStop}%
\bibitem [{\citenamefont {Henwood}\ and\ \citenamefont
  {Carey}(2007)}]{henwood2007ab}%
  \BibitemOpen
  \bibfield  {author} {\bibinfo {author} {\bibfnamefont {D.}~\bibnamefont
  {Henwood}}\ and\ \bibinfo {author} {\bibfnamefont {J.~D.}\ \bibnamefont
  {Carey}},\ }\href@noop {} {\bibfield  {journal} {\bibinfo  {journal}
  {Physical Review B}\ }\textbf {\bibinfo {volume} {75}},\ \bibinfo {pages}
  {245413} (\bibinfo {year} {2007})}\BibitemShut {NoStop}%
\bibitem [{\citenamefont {Gonz{\'a}lez-Herrero}\ \emph
  {et~al.}(2019)\citenamefont {Gonz{\'a}lez-Herrero}, \citenamefont
  {Cort{\'e}s-del R{\'\i}o}, \citenamefont {Mallet}, \citenamefont {Veuillen},
  \citenamefont {Palacios}, \citenamefont {G{\'o}mez-Rodr{\'\i}guez},
  \citenamefont {Brihuega},\ and\ \citenamefont
  {Yndur{\'a}in}}]{gonzalez2019hydrogen}%
  \BibitemOpen
  \bibfield  {author} {\bibinfo {author} {\bibfnamefont {H.}~\bibnamefont
  {Gonz{\'a}lez-Herrero}}, \bibinfo {author} {\bibfnamefont {E.}~\bibnamefont
  {Cort{\'e}s-del R{\'\i}o}}, \bibinfo {author} {\bibfnamefont
  {P.}~\bibnamefont {Mallet}}, \bibinfo {author} {\bibfnamefont
  {J.}~\bibnamefont {Veuillen}}, \bibinfo {author} {\bibfnamefont
  {J.}~\bibnamefont {Palacios}}, \bibinfo {author} {\bibfnamefont
  {J.}~\bibnamefont {G{\'o}mez-Rodr{\'\i}guez}}, \bibinfo {author}
  {\bibfnamefont {I.}~\bibnamefont {Brihuega}}, \ and\ \bibinfo {author}
  {\bibfnamefont {F.}~\bibnamefont {Yndur{\'a}in}},\ }\href@noop {} {\bibfield
  {journal} {\bibinfo  {journal} {2D Materials}\ }\textbf {\bibinfo {volume}
  {6}},\ \bibinfo {pages} {021004} (\bibinfo {year} {2019})}\BibitemShut
  {NoStop}%
\bibitem [{\citenamefont {Ivanovskaya}\ \emph {et~al.}(2010)\citenamefont
  {Ivanovskaya}, \citenamefont {Zobelli}, \citenamefont {Teillet-Billy},
  \citenamefont {Rougeau}, \citenamefont {Sidis},\ and\ \citenamefont
  {Briddon}}]{ivanovskaya2010hydrogen}%
  \BibitemOpen
  \bibfield  {author} {\bibinfo {author} {\bibfnamefont {V.}~\bibnamefont
  {Ivanovskaya}}, \bibinfo {author} {\bibfnamefont {A.}~\bibnamefont
  {Zobelli}}, \bibinfo {author} {\bibfnamefont {D.}~\bibnamefont
  {Teillet-Billy}}, \bibinfo {author} {\bibfnamefont {N.}~\bibnamefont
  {Rougeau}}, \bibinfo {author} {\bibfnamefont {V.}~\bibnamefont {Sidis}}, \
  and\ \bibinfo {author} {\bibfnamefont {P.}~\bibnamefont {Briddon}},\
  }\href@noop {} {\bibfield  {journal} {\bibinfo  {journal} {The European
  Physical Journal B}\ }\textbf {\bibinfo {volume} {76}},\ \bibinfo {pages}
  {481} (\bibinfo {year} {2010})}\BibitemShut {NoStop}%
\bibitem [{Note4()}]{Note4}%
  \BibitemOpen
  \bibinfo {note} {We note, that we use the results of {\protect \it ab initio}
  calculations for hydrogenated graphene. This is appropriate because Hydrogen
  is chemically equivalent to Tritium}\BibitemShut {NoStop}%
\bibitem [{\citenamefont {Boukhvalov}(2010)}]{boukhvalov2010modeling}%
  \BibitemOpen
  \bibfield  {author} {\bibinfo {author} {\bibfnamefont {D.}~\bibnamefont
  {Boukhvalov}},\ }\href@noop {} {\bibfield  {journal} {\bibinfo  {journal}
  {Physical Chemistry Chemical Physics}\ }\textbf {\bibinfo {volume} {12}},\
  \bibinfo {pages} {15367} (\bibinfo {year} {2010})}\BibitemShut {NoStop}%
\bibitem [{Note5()}]{Note5}%
  \BibitemOpen
  \bibinfo {note} {As an example, the value of the stiffness $\kappa $ for the
  molecular tritium according to~\cite {bodine2015assessment} is $\kappa
  \approx \SI {75}{\electronvolt \per \angstrom ^2}$. This is roughly $20$
  times as large as the corresponding value for the chemisorption (see
  Table~\ref {table:fit}). This means that the energy uncertainties $\Delta E$
  in these two cases are of the same order which is in agreement with~\cite
  {bodine2015assessment}.}\BibitemShut {Stop}%
\bibitem [{\citenamefont {Faessler}\ \emph {et~al.}(2013)\citenamefont
  {Faessler}, \citenamefont {Hodak}, \citenamefont {Kovalenko},\ and\
  \citenamefont {Simkovic}}]{faessler2013search}%
  \BibitemOpen
  \bibfield  {author} {\bibinfo {author} {\bibfnamefont {A.}~\bibnamefont
  {Faessler}}, \bibinfo {author} {\bibfnamefont {R.}~\bibnamefont {Hodak}},
  \bibinfo {author} {\bibfnamefont {S.}~\bibnamefont {Kovalenko}}, \ and\
  \bibinfo {author} {\bibfnamefont {F.}~\bibnamefont {Simkovic}},\ }\href@noop
  {} {\bibfield  {journal} {\bibinfo  {journal} {arXiv preprint
  arXiv:1304.5632}\ } (\bibinfo {year} {2013})}\BibitemShut {NoStop}%
\end{thebibliography}%

\end{document}